\documentclass[12pt,preprint]{emulateapj}

\usepackage[cmex10]{amsmath}
\usepackage{rotating}
\usepackage{url}
\usepackage{subfigure}
\usepackage{graphicx}
\usepackage{times}
\usepackage{apjfonts}
\usepackage{isotope}
\usepackage{longtable}
\usepackage{rotating}

\shorttitle{The Spectrum of Thorium from $250$ nm to $5~000$ nm}
\shortauthors{Redman et al.}

\newcommand{\nlines}{$19\,874$\space}	

\newcommand{\nblend}{$71$\space}		
\newcommand{\nambig}{$227$\space}		
\newcommand{\newlevels}{$102$\space}	
\newcommand{\newThI}{$93$\space}		
\newcommand{\newThII}{$9$\space}		
\newcommand{\nneon}{$69$\space}		

\begin{document}

\title{The Spectrum of Thorium from $250$ nm~to $5~500$ nm:
Ritz Wavelengths and Optimized Energy Levels}

\author{Stephen L. Redman\altaffilmark{1}, Gillian Nave\altaffilmark{1}, Craig J. Sansonetti\altaffilmark{1}}

\altaffiltext{1}{National Institute of Standards and Technology, Gaithersburg, MD 20899, USA}

\begin{abstract}
We have made precise observations of a thorium-argon hollow cathode lamp
emission spectrum in the region between $350$~nm and $1175$~nm using a
high-resolution Fourier transform spectrometer.  Our measurements are combined
with results from seven previously published thorium line lists
\citep{giacchetti1974proposed, zalubas1974energy, 1976Zalubas,
1983ats..book.....P, 2003JQSRT..78....1E, 2007A&A...468.1115L, kerber2008th} to
re-optimize the energy levels of neutral, singly-, and doubly-ionized thorium
(Th~I, Th~II, and Th~III).  Using the optimized level values, we calculate
accurate Ritz wavelengths for \nlines thorium lines between $250$ nm~and $5500$
nm ($40\,000$ cm$^{-1}$ to $1800$ cm$^{-1}$).  We have also found \newlevels
new thorium energy levels.  A systematic analysis of previous measurements in
light of our new results allows us to identify and propose corrections for
systematic errors in \citet{1983ats..book.....P} and typographical errors and
incorrect classifications in \citet{kerber2008th}.  We also found a large
scatter in the thorium line list of \citet{2007A&A...468.1115L}.  We anticipate
that our Ritz wavelengths will lead to improved measurement accuracy for
current and future spectrographs that make use of thorium-argon or thorium-neon
lamps as calibration standards.
\end{abstract}


\section{Introduction}

Fifty years ago the spectrum of thorium was studied in several laboratories as
a source of wavelength standards for high-resolution grating spectrographs,
resulting ultimately in an extensive list of proposed secondary standards
\citep{1970JOSA...60..474G}.  More recently, measurements with very high
precision and internal consistency were published by
\citet{1983ats..book.....P} (hereafter, PE83).  These measurements were made on
the $1$-m Fourier transform spectrometer (FTS) of the National Solar
Observatory at Kitt Peak \citep{1976JOSA...66.1081B} using a commercial
thorium-neon hollow-cathode lamp (HCL) run at the unusually high current of
$75$ mA.  The lamp was observed in five overlapping spectral regions covering
$277.7$ nm~through $1350.0$ nm ($7400$ cm$^{-1}$ through $36\,000$ cm$^{-1}$).
PE83 estimated their uncertainties ($\delta_{\sigma, \textrm{PE83}}$) in units
of 0.001 cm$^{-1}$ based upon an empirical fit to the equation

\begin{equation}
\delta_{\sigma, \textrm{PE83}} [10^{-3} cm^{-1}] = \frac{a}{I} + b\sigma
\label{eqn:pe_unc}
\end{equation}

\noindent where {\it I} is the intensity of the line, $\sigma$ is its
wavenumber ($\sigma$~[cm$^{-1}$]~$=~10^7/\lambda~\textrm{[nm]}$), and {\it a}
and {\it b} are coefficients, which they determined to be $0.5$ and
$5\times10^{-5}$ respectively (note that PE83 incorrectly reported the
coefficient {\it b} as $0.5\times10^{-5}$).  The first term in this empirical
formula is the statistical uncertainty of the measurement of the line, and the
second is the systematic uncertainty attributable to the global calibration of
the spectra.  

The thorium spectrum has been measured several more times since $1983$.  Most
notably, it was measured in the near ultraviolet (NUV) and visible regions by
\citet{2007A&A...468.1115L} (hereafter, LP07) using the High-Accuracy
Radial-velocity Planet Searcher (HARPS) \citep{2000SPIE.4008..582P}.  LP07 used
the PE83 wavelength measurements to determine the dispersion solution for HARPS
and measure faint thorium lines that did not appear in PE83.  They also
reported reduced uncertainties for many of the thorium lines observed in PE83.
The thorium spectrum was also measured in the near infrared (NIR) through
$5500$ nm ($1800$ cm$^{-1}$) at a lamp current of $320$ mA by
\citet{2003JQSRT..78....1E} (hereafter, EHW03) using the McMath $1$-m FTS.
Finally, the NIR spectrum was measured with the National Institute of Standards
and Technology (NIST) $2$-m FTS \citep{nave3progress} from $691$ nm to $5804$
nm with a hollow cathode lamp running at $20$ mA by \citet{kerber2008th}
(hereafter, KNS08). Also since 1983, a number of thorium lines have been
measured with high accuracy by laser spectroscopy \citep[ collectively referred
to hereafter as  LS3]{1984JOSAB...1..361S, 2002JOSAB..19.1711D,
DeGraffenreid:12}. These lines can be used to recalibrate the results of PE83,
reducing the systematic contribution to their uncertainties.

Thorium-argon (Th/Ar) HCLs are installed on many of the highest-precision
astronomical spectrographs, including the European Southern Observatory's
(ESO's) CRyogenic high-resolution InfraRed Echelle Spectrograph (CRIRES)
\citep{kaeufl2004crires}, the High-Resolution Spectrograph (HRS)
\citep{1998SPIE.3355..387T} on the Hobby-Eberly Telescope (HET)
\citep{1998SPIE.3352...34R}, and HARPS.  HARPS is currently the most precise
astronomical spectrograph in the world, and typically achieves radial velocity
precision below $1$ m/s (a relative precision of $3$ parts in $10^9$)
\citep{2009A&A...507..487M}.

Exoplanet radial velocity measurements, such as those made with the HARPS
spectrograph, attempt to detect periodic variations in stellar wavelengths;
therefore, they do not usually depend upon the absolute accuracy of the
calibration source as long as the calibration lines are precisely reproducible.
However, systematic errors in the spectrograph can increase the uncertainty of
the measured radial velocities. An example of such systematic errors can be
seen in figure $4$ of \citet{2010MNRAS.405L..16W}, where the authors show the
difference between the dispersion solutions found using a thorium-argon lamp
and a laser frequency comb across the same order of the HARPS spectrograph.
Since the LP07 line list was based on observations with the HARPS spectrograph,
such errors in the dispersion solution undoubtedly affect their thorium
wavelengths.

In an effort to provide the physics and astrophysics communities with thorium
wavelengths of higher accuracy, we have re-measured more than $1600$ thorium
lines in the NUV to NIR.  Additionally, we have performed least-squares energy
level optimizations for Th~I, Th~II, and Th~III using classified lines from our
measurements and previous work.  The optimization provides precise level values
from which we calculate Ritz wavelengths for all observed lines of thorium that
we can classify throughout the optical and NIR.  Since these Ritz wavelengths
are calculated from globally optimized energy levels, they are usually more
precise than the measured wavelengths.

In \S~\ref{sec:motivation}, we lay out the motivation for this paper.  In
\S~\ref{sec:line_list_construction}, we describe our new measurements and
compare them to previously reported line lists.  In \S~\ref{sec:ritz}, we
describe how we adjusted and used data from various sources to classify thorium
lines throughout the optical and near infrared, re-optimized the levels of
neutral, singly-, and doubly-ionized thorium, and calculated Ritz wavelengths.
This section includes an abbreviated line list.  Finally, in
\S~\ref{sec:conclusions}, we summarize the work and draw some conclusions.

\section{Motivation for an Updated Line List}
\label{sec:motivation}

Atomic emission lines are caused by transitions of electrons from an upper
energy level to a lower energy level.  Atomic energy levels are usually not
measured directly, but can be inferred by searching large numbers of emission
or absorption lines for common differences between unclassified lines and known
energy levels.  If the energies of the combining levels are accurately known,
the wavenumber and corresponding wavelength of the transition can be calculated
by simply differencing the energies.  These calculated values are known as Ritz
wavelengths and are often more accurate than experimental measurements due to
the large number of measured transitions that determine each energy level.

In the case of thorium, a large number of energy levels of Th~I and Th~II were
accurately determined in the work of PE83.  Ritz wavelengths based on these
energy levels can be used to classify observed lines in newly acquired thorium
spectra.  Observed spectral lines that do not match Ritz wavelengths within
their expected uncertainties fall into several categories.  Assuming that the
spectrograph itself is not responsible for introducing systematic errors,
either (a) the spectral line is formed by a transition between two energy
levels one or both of which is not known, (b) one or both energy level values
is inaccurate, (c) the observed spectral line is attributable to some other
element contaminating the source, or (d) the observed line is shifted by a
blend with another line, which is not resolved in the spectrum.

This work was undertaken in an attempt to determine whether some of the many
unidentified lines in the LP07 line list could be classified by comparing their
measured wavelengths to thorium Ritz wavelengths.  We hoped that this
comparison would provide two distinct benefits. First, it would help us
determine which unidentified lines in the LP07 line list are thorium lines that
can be used for precise wavelength calibration and which ones are either argon
lines (which are affected by lamp variations at the $0.003$ nm level --- tens
of m/s) \citep{2006SPIE.6269E..23L} or unknown lines (which also should not be
used for calibration).  Second, it would allow us to propose accurate Ritz
wavelengths for the entire set of classified lines of LP07.  Unfortunately, we
found it generally impossible to definitively match the experimental results of
LP07 with the Ritz wavelengths.

\begin{figure}[htbp]
\includegraphics[width=0.5\textwidth]{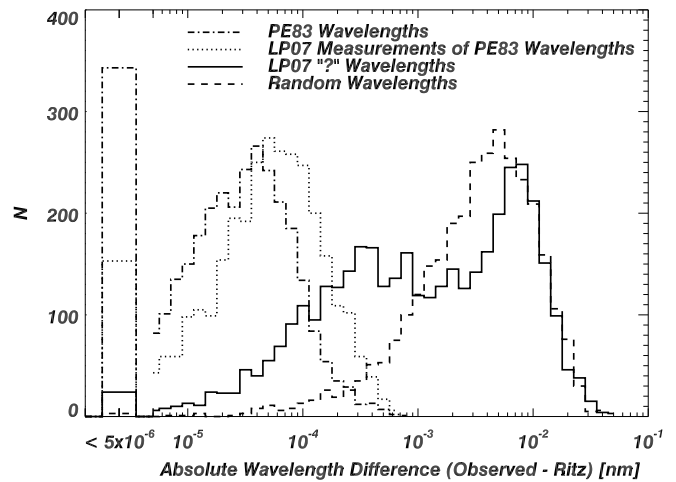}
\caption{The distribution in differences between the measured wavelengths and
closest Ritz wavelengths of four populations of thorium wavelength
measurements.  All Ritz wavelengths were calculated using energy levels
published in the online actinides database
(http://web2.lac.u-psud.fr/lac/Database/Tab-energy/Thorium/Th-el-dir.html).  The thorium
lines of \citet{1983ats..book.....P} are shown as a dash-dotted line, the
unidentified emission lines of \citet{2007A&A...468.1115L} are shown as a solid
line, and the PE83 thorium lines as measured by LP07 are shown as a dotted
line.  A set of uniformly-distributed random wavelengths are shown with the
dashed histogram.  The thorium lines of PE83 show the best agreement with the
Ritz wavelengths.}
\label{figure1}
\end{figure}

Figure~\ref{figure1} shows the distribution of the absolute differences between
Ritz wavelengths and the closest measured thorium lines of PE83 (dash-dotted
line) and the same set of lines as measured by LP07 (dotted line).  The energy
levels were taken from an online database of published and unpublished actinide
energy
levels\footnote{\url{http://web2.lac.u-psud.fr/lac/Database/Tab-energy/Thorium/Th-el-dir.html}},
with uncertainties of $0.001$ cm$^{-1}$ for Th~I and Th~II levels.  For
clarity, the distributions have been truncated below $5\times10^{-6}$.  A
single histogram bar below this value indicates the sum of the distribution
tail below this value.  The standard deviation of the difference between PE83
and the Ritz wavelengths is $8.2\times10^{-5}$ nm, while the corresponding
quantity for LP07 is $13.2\times10^{-5}$ nm.  The black histogram (solid line)
is the corresponding distribution for the many unidentified emission lines
reported by LP07.  This histogram is double-peaked, probably due to the
presence of both (a) thorium lines that match their Ritz wavelengths to within
about $0.001$ nm (lower-difference peak) and (b) argon lines, unknown lines,
and thorium transitions between unknown energy levels (higher-deviation peak).
This latter peak is similar to the distribution found when random wavelengths
are compared to the $21\,440$ theoretically-possible thorium Ritz wavelengths
between $370$ and $700$ nm ($14\,000$ cm$^{-1}$ and $27\,000$ cm$^{-1}$) as
illustrated by the dashed line in Figure~\ref{figure1}.

It is clear that the LP07 lines do not agree with the Ritz wavelengths as well
as the same lines observed by PE83.  This is to be expected, since many of the
energy levels come from the PE83 data.  However, the disagreement between the
unidentified LP07 lines and the closest Ritz wavelengths (solid histogram) is
surprisingly large.  Surely many of these lines must involve transitions
between known thorium energy levels, but the distribution does not resemble the
LP07 measurements of PE83 wavelengths (dotted histogram).  There are two
possible explanations for this observation.  First, it is possible that the
energy levels of Th~I and Th~II as presented in PE83 need to be updated.  A
re-optimization of the energy levels using the wavenumbers of LP07 might
provide a more accurate and internally consistent set of energies than those
reported by PE83.

The second possibility is that the LP07 wavelengths are less accurate than the
PE83 wavelengths.  This would be surprising, since one of the primary
motivations behind the LP07 line list was to provide lower uncertainties than
PE83.  Their uncertainties, calculated by using Equation~\ref{eqn:pe_unc}, are
typically between $2\times10^{-5}$ nm and $10\times10^{-5}$ nm ($15$ m/s to
$62$ m/s).  By comparison, the LP07 uncertainties for the same lines (as
published) are between $1\times10^{-5}$ nm and $76\times10^{-5}$ nm ($7$ m/s to
$600$ m/s), with several thousand lines having uncertainties substantially
reduced with respect to PE83.

In an attempt to understand the discrepancies between the results of PE83 and
LP07, we recorded a new thorium spectrum with the $2$-m FTS at NIST.  There are
several advantages to using an FTS (as opposed to an echelle spectrograph) to
measure wavelengths in emission spectra.  First and foremost, the dispersion
solution of an FTS is linear in wavenumber.  As long as a few standard lines,
measured independently to high accuracy, are in the observed spectrum, the
entire dispersion solution can be calibrated to a similar accuracy.  Second,
the signal-to-noise ratio (S/N) in the spectrum can be improved by increasing
the integration time without fear of saturating even the brightest emission
lines.  Third, the very high resolving power of the FTS (as high as $10^7$ in
the case of the NIST $2$-m FTS) means that blended lines are less common and
that noble gas lines can often be distinguished from heavier elements on the
basis of their measured widths alone.  Finally, the high resolution of the FTS
means that uncertainties in determining the positions of individual lines are
much smaller than can be achieved with lower-resolution instruments (see
Eqn.~\ref{eqn:unc}).

\section{New Measurements of Thorium}
\label{sec:line_list_construction}

\subsection{Experimental Setup}

We observed the spectrum of a commercial sealed Th/Ar HCL running at $25$~mA
for almost $15$ hours starting on $02$ November $2011$.  The spectrum was
recorded between $0$ cm$^{-1}$ and $30\,000$ cm$^{-1}$ at a sampling resolution
of $0.02$ cm$^{-1}$.  The region of observation was limited to between
$8500$~cm$^{-1}$ and $28\,000$~cm$^{-1}$ ($360$ nm and $1200$ nm) by the
sensitivity of the silicon photodiode detector.  The FTS enclosure was
evacuated to a  pressure of about $2.5$~Pa (0.019~Torr) throughout the
observation.  Low resolution spectra of a radiometrically calibrated tungsten
ribbon lamp were recorded before and after the thorium spectrum so that we
could determine the spectral response of the FTS.  The response function
derived from these spectra was used to obtain radiometrically calibrated
relative intensities for the emission lines from the HCL.

Our new spectrum was calibrated with respect to thorium lines precisely
measured by  LS3.  It provides a direct comparison to the results of PE83
that spans portions of all five of their thorium-neon spectra.  Because Palmer
and Engleman did not have good internal standards across the entire spectrum,
they arbitrarily chose to take as their calibration factor the finite aperture
correction for the $11\,000$ cm$^{-1}$ to 18\,000 cm$^{-1}$ ($550$ nm to $900$
nm) spectral region. This calibration was propagated to the other four spectral
regions by using overlapping lines.  No internal standard lines were used for
calibration. Thus it is very valuable to have a comparison spectrum with a
well-defined calibration that overlaps with all of the PE83 regions.

\subsection{Measurements and Uncertainties}
\label{ssec:measurements}

The spectrum was measured by fitting a Voigt function to each emission line
with a S/N of greater than $20$ using the program XGremlin
\citep{nave3progress}, an X-windows implementation of {\sc gremlin}
\citep{brault1989high}.  The spectrum was interpolated by convolution with the
FTS instrumental function before fitting the Voigt profile.  The initial
wavenumber scale was based on the optical path difference measured by the
helium-neon control laser of the FTS as the interferogram was recorded.
However, even in the case of perfect alignment, the optical path difference for
an emission source is slightly different from that measured by the control
laser because of the finite instrumental aperture.  This results in a
stretching of the wavenumber scale that we correct in the usual way by
determining a multiplicative correction factor $\kappa_i$ from internal
standard lines:

\begin{equation}
\kappa_{\textrm{i}} = \frac{\sigma_{\textrm{std,i}}}{\sigma_{\textrm{obs,i}}} - 1
\label{eqn:wncf}
\end{equation}

\noindent where $\sigma_{\textrm{std,i}}$ is the accurately known wavenumber of
the standard line, and $\sigma_{\textrm{obs,i}}$ is the uncorrected wavenumber
of the same line.  The uncertainty in the individual wavenumber correction
factor ($\delta_{\kappa\textrm{,i}}$) is the sum in quadrature of the relative
uncertainty in the standard wavenumber and the relative statistical uncertainty
of the measured line:

\begin{equation}
\delta_{\kappa\textrm{,i}} = \sqrt{ \left(\frac{\delta_{\textrm{std,i}}}{\sigma_{\textrm{std,i}}}\right)^2 +
						  \left(\frac{\delta_{lc,i}}{\sigma_{\textrm{obs,i}}}\right)^2 }
\end{equation}

\noindent where $\delta_{lc,i}$ is the statistical uncertainty in the line
centroid derived from equation $9.2$ of \citet{davis9fourier}.  Our spectra are
under-sampled, so there is only one statistically-independent point in a line
width:

\begin{equation}
\delta_{lc,i} = \frac{W_{i}}{S/N_{i}\sqrt{N_{w,i}}}
\label{eqn:lc_unc}
\end{equation}

\noindent where $W_{i}$ is the measured line width at half-maximum, $S/N_{i}$
is the measured signal-to-noise ratio of the line, and $N_{w,i}$ is the number
of significant points in the width of the line, which is equal to the line
width divided by the spectral resolution.

The overall correction factor for the spectrum, $\kappa$, is taken to be the
mean of the individual line correction factors, weighted by the inverse square
of their uncertainties.

\begin{equation}
\kappa = \frac{\sum \kappa_{\textrm{i}} \delta_{\kappa\textrm{,i}}^{-2}}{\sum{\delta_{\kappa\textrm{,i}}^{-2}}}
\label{eq:kappa}
\end{equation}

\noindent All wavenumbers in the spectrum are corrected by multiplying by the
factor ($1 + \kappa$).  The uncertainty ($\delta_\kappa$) in $\kappa$ is taken
to be the standard error of the mean of individual $\kappa_i$ values.

The final uncertainty of each of our measured wavenumbers is the sum in
quadrature of the calibration uncertainty and the statistical uncertainty of
the line centroid:

\begin{equation}
\delta_{\sigma,i} = \sqrt{\delta_{\kappa}^2 \sigma_i^2 + \delta_{lc,i}^2}
\label{eqn:unc}
\end{equation}

\begin{figure}[htbp]
\includegraphics[height=.5\textwidth,angle=90.]{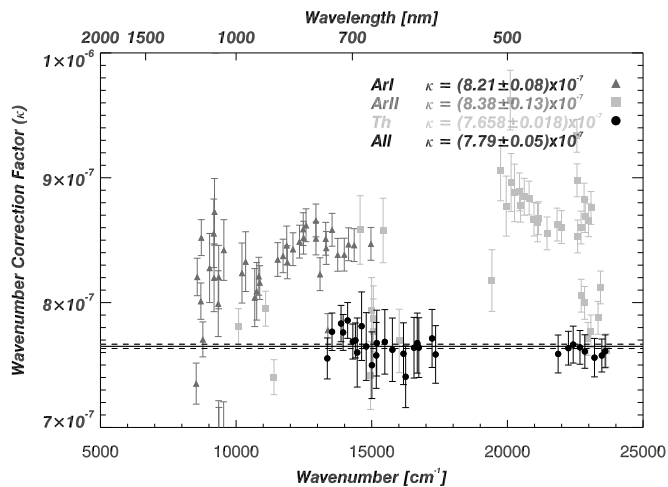}
\caption{The wavenumber correction factor $\kappa$ determined for the Th/Ar
spectrum obtained at NIST.  We compared three different standards to our
measured wavenumbers: neutral argon lines, low-energy Ar-II lines, and 30
thorium lines that have been measured to precisions of a few parts in $10^8$ by
 LS3.}
\label{figure2}
\end{figure}

The determination of the wavenumber correction factor for our Th/Ar spectrum is
illustrated in Figure~\ref{figure2}.  We compared values of the correction
factor as determined from three atomic species: neutral argon (Ar~I),
singly-ionized argon (Ar~II), and neutral thorium (Th~I).  The Ar~I standards
are low-excitation lines from \citet{whaling2002argon} with corrections from
\citet{sansonetti2007comment}, while the Ar~II standards are from
\citet{whaling1995argon}.  For thorium standards, we used $30$ of the $31$
lines measured by laser spectroscopy by LS3.  One line (at $15~966.406~1$
cm$^{-1}$, $626.315\,022$ nm) was not used because it is suspected to be a
blend \citep{1984JOSAB...1..361S}, and indeed it was an outlier compared to our
newest measurements at NIST.  We suspect that this line is a blend of the
transition between the Th~I energy levels $10\,414_4$ and $26\,380_5$
($15\,966.405\,0$ cm$^{-1}$, $626.315\,066$ nm) and the transition between the
Th~I energy levels $13\,175_4$ and $29\,141_5$ ($15\,966.416\,9$ cm$^{-1}$,
$626.314\,600$ nm).  The weighted mean of the wavenumber correction factor
($\kappa$, Equation~\ref{eq:kappa}) and the standard error of the mean of the
wavenumber correction factor ($\delta\kappa$) for the different standards are
reported in the legend of Figure~\ref{figure2}.

From Figure~\ref{figure2}, it is immediately clear that the Th~I, Ar~I, and
Ar~II standards do not agree with each other.  It is likely that this
disagreement results from a difference in the mean optical path through the FTS
for thorium and argon emission.  Since the thorium lines arise almost
exclusively from the interior of the cathode, while the argon emission is
broadly distributed in the gas between the cathode and anode, any imperfection
in the alignment of the cathode with the optical axis of the spectrograph will
produce a systematic shift between these sources.  For our purposes, this
disagreement is unimportant, since we are only concerned with obtaining
accurate wavenumbers for thorium.  We therefore take $\kappa = (7.649 \pm
0.017)\times10^{-7}$, derived from the thorium lines only, as the correction
factor for the spectrum.

\subsection{Results}
\label{sec:results}

We measured the wavenumbers of $1644$ thorium lines between $350$ nm ($28\,570$
cm$^{-1}$) and $1175$ nm ($8510$ cm$^{-1}$).  Identifications of lines as
thorium were based primarily on their observed line widths.  Thorium line
widths were typically between $0.03$ cm$^{-1}$ and $0.04$ cm$^{-1}$
($1\times10^{-6}$ nm and $3\times10^{-6}$ nm), and argon line widths were
typically between $0.05$ cm$^{-1}$ and $0.09$ cm$^{-1}$ ($3\times10^{-6}$ nm).
Thorium lines that could be matched unambiguously to Ritz wavelengths have been
classified.  A sample of our new measurements is presented in
Table~\ref{table1}, which is generally self explanatory.  The rightmost column
contains the difference between our measured wavenumber and the Ritz wavenumber
calculated from our re-optimized thorium energy levels as discussed below.

\begin{table*}
\caption{New Thorium Measurements (RNS13)}
\centering
\begin{tabular}{llllcccccccr}
\hline \hline
Wavenumber			&	Wavenumber				&	Vacuum			&	Wavelength			&	Line Width			&	Measured	&	Species	&	Odd			&	Odd		&	Even			&	Even		&	$\sigma - \sigma_{Ritz}$	\\
				&	Uncertainty				&	Wavelength			&	Uncertainty			&	(W)				&	S/N		&			&	Energy		&	J		&	Energy		&	J		&						\\
($\sigma$) [cm$^{-1}$]	&	($\delta_\sigma$) [cm$^{-1}$]	&	($\lambda$) [nm]		&	($\delta_\lambda$) [nm]	&	[$10^{-3}$cm$^{-1}$]	&			&			&	[cm$^{-1}$]		&			&	[cm$^{-1}$]		&			&	[cm$^{-1}$]			\\
\hline

$  11446.3345  $	&	  $      0.0008  $	&	  $   873.64213  $	&	  $     0.00006  $	&	  $ 22  $	&	  $      26  $	&	 Th I	&	  $   20566  $	&	  $           4  $	&	  $   32012  $	&	  $           4  $	&	  $      0.0004$   \\
$ 11448.42845  $	&	  $     0.00023  $	&	  $  873.482334  $	&	  $    0.000017  $	&	  $ 23  $	&	  $      96  $	&	 Th I	&	  $   23049  $	&	  $           1  $	&	  $   11601  $	&	  $           1  $	&	  $     0.00010$   \\
$  11458.9925  $	&	  $      0.0012  $	&	  $   872.67707  $	&	  $     0.00009  $	&	  $ 25  $	&	  $      18  $	&	 Th I	&	  $   25306  $	&	  $           2  $	&	  $   13847  $	&	  $           2  $	&	  $      0.0013$   \\
$  11459.8558  $	&	  $      0.0016  $	&	  $   872.61133  $	&	  $     0.00013  $	&	  $ 22  $	&	  $      12  $	&	 Th I	&	  $   13945  $	&	  $           3  $	&	  $   25405  $	&	  $           4  $	&	  $      0.0008$   \\
$  11462.5573  $	&	  $      0.0020  $	&	  $   872.40567  $	&	  $     0.00015  $	&	  $ 23  $	&	  $      10  $	&	Th II	&	  $   27249  $	&	  $         7/2  $	&	  $   15786  $	&	  $         5/2  $	&	  $     -0.0020$   \\
$   11465.228  $	&	  $       0.003  $	&	  $   872.20247  $	&	  $     0.00022  $	&	  $ 28  $	&	  $      8.  $	&	Th II	&	  $   26770  $	&	  $        11/2  $	&	  $   15305  $	&	  $         9/2  $	&	  $      -0.000$   \\
$   11472.283  $	&	  $       0.003  $	&	  $   871.66611  $	&	  $     0.00021  $	&	  $ 27  $	&	  $      8.  $	&	 Th I	&	  $   24769  $	&	  $           3  $	&	  $   13297  $	&	  $           4  $	&	  $       0.000$   \\
$  11473.0895  $	&	  $      0.0017  $	&	  $   871.60481  $	&	  $     0.00013  $	&	  $ 23  $	&	  $      12  $	&	 Th I	&	  $   24561  $	&	  $           3  $	&	  $   13088  $	&	  $           3  $	&	  $     -0.0011$   \\
$  11477.3570  $	&	  $      0.0012  $	&	  $   871.28073  $	&	  $     0.00009  $	&	  $ 21  $	&	  $      17  $	&	 Th I	&	  $   19588  $	&	  $           5  $	&	  $   08111  $	&	  $           4  $	&	  $     -0.0007$   \\
$ 11478.91339  $	&	  $     0.00008  $	&	  $  871.162597  $	&	  $    0.000006  $	&	  $ 22  $	&	  $      27  $	&	 Th I	&	  $   15166  $	&	  $           3  $	&	  $   03687  $	&	  $           2  $	&	  $    -0.00000$   \\
$  11481.3855  $	&	  $      0.0011  $	&	  $   870.97502  $	&	  $     0.00008  $	&	  $ 23  $	&	  $      20  $	&	 Th I	&	  $   15490  $	&	  $           5  $	&	  $   26971  $	&	  $           4  $	&	  $      0.0002$   \\
$  11484.6810  $	&	  $      0.0013  $	&	  $   870.72510  $	&	  $     0.00010  $	&	  $ 24  $	&	  $      17  $	&	 Th I	&	  $   17847  $	&	  $           2  $	&	  $   06362  $	&	  $           2  $	&	  $     -0.0005$   \\
$  11486.2093  $	&	  $      0.0012  $	&	  $   870.60924  $	&	  $     0.00009  $	&	  $ 21  $	&	  $      16  $	&	 Th I	&	  $   25690  $	&	  $           5  $	&	  $   14204  $	&	  $           5  $	&	  $      0.0006$   \\
$  11492.9482  $	&	  $      0.0021  $	&	  $   870.09876  $	&	  $     0.00016  $	&	  $ 32  $	&	  $      11  $	&	 Th I	&	  $   23093  $	&	  $           2  $	&	  $   11601  $	&	  $           1  $	&	  $     -0.0029$   \\
$  11507.1669  $	&	  $      0.0007  $	&	  $   869.02364  $	&	  $     0.00005  $	&	  $ 23  $	&	  $      32  $	&	 Th I	&	  $   15490  $	&	  $           5  $	&	  $   26997  $	&	  $           6  $	&	  $     -0.0030$   \\
$  11509.4902  $	&	  $      0.0024  $	&	  $   868.84821  $	&	  $     0.00018  $	&	  $ 20  $	&	  $      8.  $	&	Th II	&	  $   17722  $	&	  $         9/2  $	&	  $   06213  $	&	  $         9/2  $	&	  $     -0.0049$   \\

\hline
\end{tabular}

A sample of the new thorium measurements made with the NIST 2-m FTS.
Table~\ref{table1} is published in its entirety in the electronic edition of
ApJSS.  This table includes unclassified lines that do not appear in the
compiled thorium line list (Table~\ref{table6}).
\label{table1}
\end{table*}

\subsection{Comparison with Previous Thorium Line Lists}
\label{sec:comparison}

We compared our thorium measurements to three previously published line lists:
the Th/Ne measurements of PE83, which were made using the Kitt Peak FTS; the
Th/Ar measurements of KNS08, which were made using the NIST 2-m FTS; and the
Th/Ar measurements of LP07, which were made using the HARPS echelle
spectrograph.  The Th/Ar spectrum of EHW03 includes lines between $1798$
cm$^{-1}$ and $9180$ cm$^{-1}$ ($1089$ nm to $5562$ nm), mostly outside of our
observed range.  The line list of EHW03 is in good agreement with the
measurements of KNS08 in their region of overlap.

It is not uncommon for papers in the scientific literature to report general
uncertainties, rather than uncertainties for individual lines. The relative
contribution of systematic and random factors to these general uncertainties is
often ill-defined. In such cases, it is difficult or impossible to assign
appropriate uncertainties to the wavelengths if the data are recalibrated using
improved standards that reveal previously-unknown systematic errors. We have
estimated the statistical uncertainties of previously-reported measurements by
comparing the measured wavenumbers to the Ritz wavenumbers (after optimizing
the energy levels; see Section 4). This process conflates the statistical
uncertainties with any systematic uncertainties, so in these cases, we added
the statistical and systematic uncertainties linearly. We found that the
statistical uncertainties could be represented by functions of the form:

\begin{equation}
\delta_{\sigma,i} = \delta_{\kappa}\sigma_i + c / I_i^d
\label{eqn:unc2}
\end{equation}

\noindent where $c$ and $d$ are coefficients, and $I_i$ is the reported
intensity of the line.  This function is similar to the equation used by PE83
to estimate their statistical uncertainties (see Eqn.~\ref{eqn:pe_unc}).

\subsubsection{Comparison with Palmer \& Engelman (1983)}
\label{ssec:comparison_pe83}

The measurements of PE83 were made by combining results from five overlapping
spectra from the Kitt Peak FTS.  Until the introduction of LP07, this line list
was widely used throughout the astrophysical community to calibrate Th/Ar
spectra.

\begin{figure}[htbp]
\includegraphics[height=.5\textwidth,angle=90.]{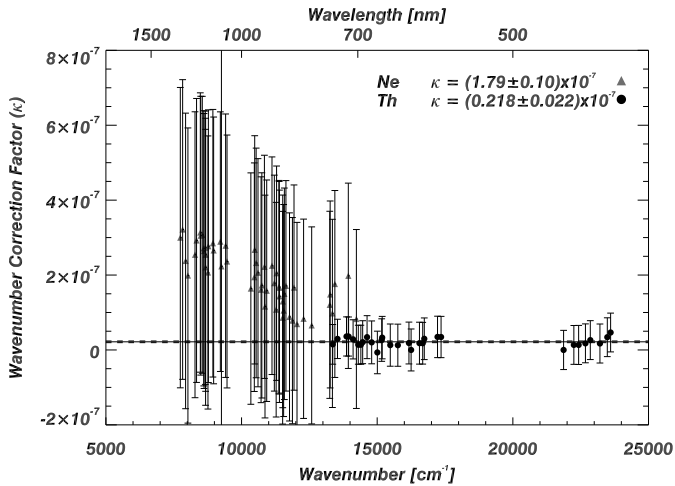}
\caption{The wavenumber correction factor $\kappa$ determined for the Th/Ne
line list of PE83, using their published wavenumbers and wavenumber
uncertainties.  This wavenumber correction factor is in addition to the
wavenumber correction factor of $5.05838\times10^{-7}$ applied to the PE83
spectrum by its authors.}
\label{figure3}
\end{figure}

In $1983$, Palmer and Engleman had no internal standards known with sufficient
accuracy to determine their wavenumber correction factor.  Since then, the
wavenumbers of several thorium lines have been measured via laser spectroscopy
by \citet{1984JOSAB...1..361S} (fifteen lines), \citet{2002JOSAB..19.1711D}
(seven lines), and \citet{DeGraffenreid:12} (eight lines).
\citet{saloman2004wavelengths} also compiled thousands of neon lines, some of
which are recommended as standards.  Using the thorium and neon wavenumber
measurements of PE83, we have redetermined the wavenumber correction factor of
the PE83 line list (Figure~\ref{figure3}).  In light of the trend and large
scatter in the neon results, we have used only the thorium lines to calculate a
revised wavenumber correction factor for the results of PE83.  Note that a
similar trend in the neon standards was seen in \citet{2011ApJS..195...24R}.  

The PE83 line list is a compilation of several, independently-measured spectra.
It is unlikely that they all share the same uncertainty relationship.
Therefore, we have estimated the uncertainties of PE83 in each of the spectral
regions by fitting the standard deviation of the differences between the
measured wavenumber and the Ritz wavenumbers as a function of intensity.  The
uncertainties are then given by Equation~\ref{eqn:unc2}.  The $c$ and $d$
coefficients are different for the different spectral regions of the line list
(which was segmented at --- roughly --- $11\,000$ cm$^{-1}$, $18\,000$
cm$^{-1}$, $21\,000$ cm$^{-1}$, $25\,000$ cm$^{-1}$, and $30\,000$ cm$^{-1}$,
or $900$ nm, $550$ nm, $480$ nm, $400$ nm, $330$ nm), and are summarized in
Table~\ref{table2}.  We set the calibration uncertainty throughout to be
$1.2\times10^{-8}\times\sigma_{PE83}$.

\begin{table}
\caption{Uncertainty Coefficients for PE83 (Eqn.~\ref{eqn:unc2})}
\centering
\begin{tabular}{lcc}
\hline \hline
Spectral Range		&	c			&	d	\\
$[$cm$^{-1}$]			&	[cm$^{-1}$]	&		\\
\hline
$< 11\,000$			&	$0.00053$		&	$0.678$	\\	
$11\,000$ --- $18\,000$	&	$0.00192$		&	$0.482$	\\
$18\,000$ --- $21\,000$	&	$0.00165$		&	$0.462$	\\
$21\,000$ --- $25\,000$	&	$0.00171$		&	$0.404$	\\
$25\,000$ --- $30\,000$	&	$0.00331$		&	$0.405$	\\
$> 30\,000$			&	$0.00324$		&	$0.531$	\\
\hline
\end{tabular}

\label{table2}
\end{table}

\begin{figure*}[htbp]
\begin{center}
\includegraphics[width=0.9\textwidth]{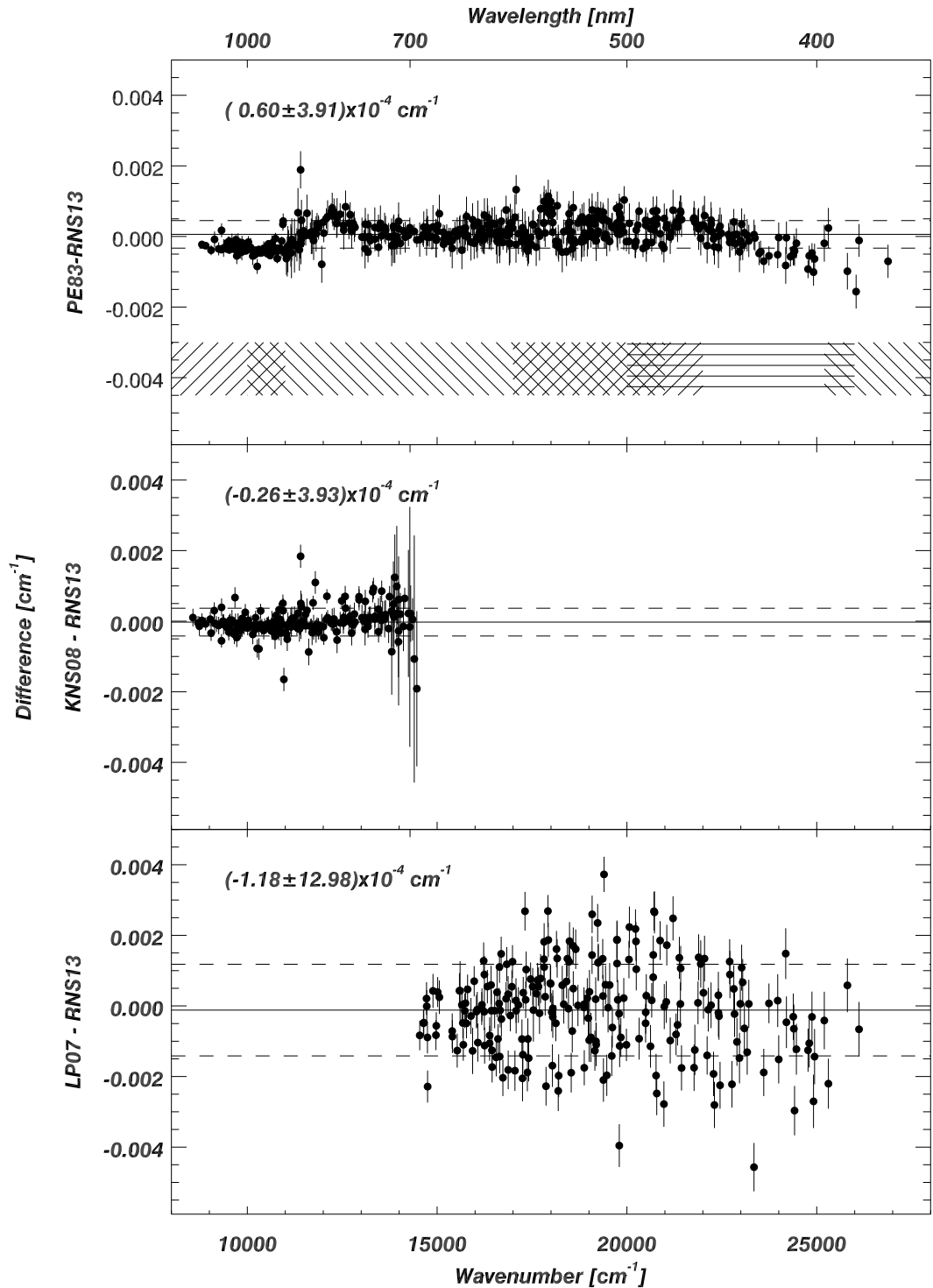}
\caption{A comparison between our measured wavenumbers and three historical
sources, all shown on the same scale: PE83 (top), KNS08 (middle), and LP07
(bottom).  These plots only compare lines for which our uncertainty was less
than $0.0004$ cm$^{-1}$ and the signal-to-noise ratio of our lines was greater
than $100$ (about $10\%$ of the lines).  The error bars are the sum in
quadrature of our NIST measurements and the uncertainties of PE83, LP07, or
KNS08 (respectively).  The mean and standard deviation of each difference is
printed in the upper-left corner of each plot.  The PE83 measurements have been
corrected by a $\kappa$ of $2.2\times10^{-8}$ (see Figure~\ref{figure3}).  The
PE83 uncertainties are the published uncertainties (using
Equation~\ref{eqn:pe_unc}). Hash marks in the lower part of this panel indicate
the overlap between different spectra in PE83.}
\label{figure4}
\end{center}
\end{figure*}

The difference between our new thorium measurements (hereafter, RNS13) and the
re-calibrated results of PE83 --- shown in the top panel of
Figure~\ref{figure4} --- reveal some minor systematic trends in the data of
PE83.  First, there are at least two places in the spectrum where the
differences systematically rise and fall.  These can be seen at about $12\,000$
cm$^{-1}$ and $18\,000$ cm$^{-1}$ ($830$ nm and $550$ nm, respectively).  The
feature at $18\,000$ cm$^{-1}$ corresponds to a region where measurements from
overlapping spectra were combined, but the feature at $12\,000$ cm$^{-1}$ is
not at a boundary point between two spectra.  The latter feature does
correspond to a local minimum in the reflectivity of aluminum, so we speculate
that PE83 used a low-resolution transform when determining the phase.  Second,
the PE83 wavenumbers are systematically smaller than ours below $11\,000$
cm$^{-1}$ ($900$ nm) and above about $20\,000$ cm$^{-1}$ ($500$ nm). The trend
above $20\,000$  cm$^{-1}$ corresponds to the linear trend in the wavenumber
correction factor of PE83 shown in Fig.~\ref{figure3}.  Since we see this same
deviation when we compare the PE83 wavenumber measurements to the
laser-spectroscopy measurements of \citet{DeGraffenreid:12}, but not when we
compare our own measurements to these standards, this suggests the trend exists
in the PE83 line list.

\subsubsection{Comparison with Kerber, Nave, \& Sansonetti (2008)}

The near-infrared Th/Ar line list of \citet{kerber2008th} was based on
observations with the 2-m FTS at NIST, the same spectrometer we used to make
our measurements.  It should be noted that we are using a different beam
splitter, mirrors, and detector, and we would not expect any systematics to be
common between the observations.  Our measurements are in agreement with
\citet{kerber2008th} to within $1$ part in $10^8$.  We see no systematic bump
in the KNS08 line list measurements around $12\,000$ cm$^{-1}$ ($830$ nm) --
indeed, KNS08 also saw a systematic deviation from the wavenumbers of PE83 (see
their figure~$7$).  This is further evidence that this deviation represents a
local systematic error in the line list of PE83.

\subsubsection{Comparison with Lovis \& Pepe (2007)}

\citet{2007A&A...468.1115L} measured the wavelengths of thorium lines between
about $380$ nm and $690$ nm ($26\,000$ cm$^{-1}$ and $14\,500$ cm$^{-1}$,
respectively) using the echelle spectrograph HARPS.  Because of the high
sensitivity of their echelle spectrograph, they saw thousands of lines not
reported by PE83.  The primary motivations of their work were to provide more
accurate thorium wavelengths than those published in PE83 and to measure the
wavelengths of unknown emission lines that were observable in the HARPS
spectra.  However, the comparison between our thorium measurements and those of
LP07 shows a large scatter, seen  in the bottom panel of Fig.~\ref{figure4},
that is about $2.5$ times larger than the average uncertainty.  This is in
sharp contrast to our agreement with the PE83 thorium line measurements (top
panel of Fig.~\ref{figure4}), and suggests that the LP07 uncertainties were
underestimated.

This large scatter probably comes from a number of sources, including
systematic offsets in the HARPS detectors \citep[such as the imperfect
stitching of the multiple detectors in the spectrograph focal plane found
by][]{2010MNRAS.405L..16W} and shifts due to unresolved blending of adjacent
lines.  LP07 assumed (incorrectly) that their measurements were much more
precise than those of PE83 because of the high S/N of their spectra.  As a
consequence, they attributed any differences between their measurements and
those of PE83 to systematic errors or random noise in the measurements of the
latter.  Our analysis suggests that the discrepancies actually come largely
from systematic errors in the measurements of LP07.\footnote{It is worth noting
that grating spectrometers have much higher efficiencies than Fourier transform
spectrometers, and represent one of the few ways to find and measure the
fainter thorium lines.  It should be possible to use grating spectrographs to
measure these fainter thorium lines, but it is essential that the FTS-measured
thorium lines be treated as standards throughout the process.  The more
accurate FTS measurements should be used to estimate (and possibly correct) the
systematic errors of the grating spectrograph dispersion solutions before
estimating the wavelengths and uncertainties of the unknown lines.}

\begin{figure}[htbp]
\includegraphics[width=0.5\textwidth]{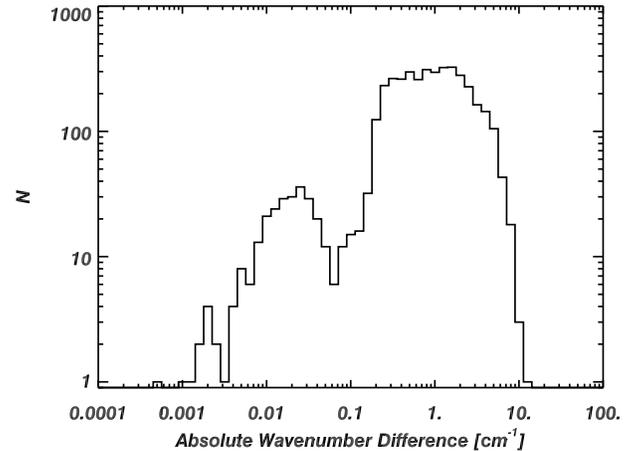}
\caption{The absolute difference between the LP07 `?' lines and the closest
known PE83 thorium lines.  The smaller bump below about $0.05$ cm$^{-1}$ made
us suspect that there was a population of thorium lines that were in PE83, but
were misidentified by LP07 as new thorium lines.  A wavenumber difference of
$0.05$ cm$^{-1}$ is a wavelength difference of $0.008$ nm at $400$ nm and
$0.025$ nm at $700$ nm.}
\label{figure5}
\end{figure}

\begin{figure}[htbp]
\includegraphics[width=0.5\textwidth]{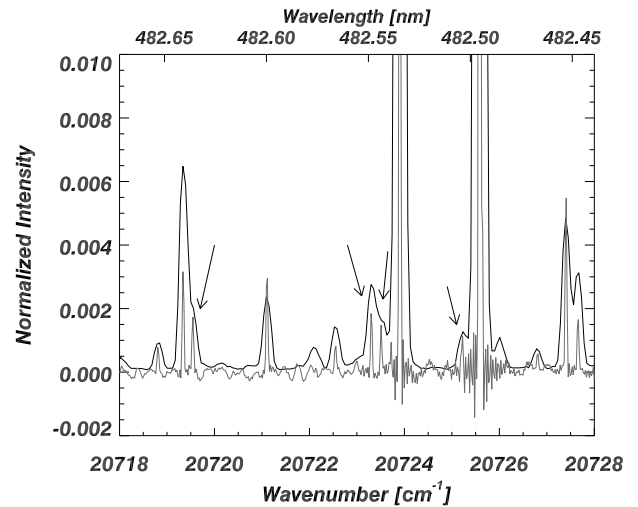}
\caption{Comparison of the Th/Ar spectrum observed with HARPS (LP07, black) and
with the Kitt Peak FTS (PE83, grey).  The four arrows indicate lines that LP07
failed to correlate with the corresponding lines measured in the higher
resolution spectra of PE83.}
\label{figure6}
\end{figure}

\begin{figure}[htbp]
\includegraphics[width=0.5\textwidth]{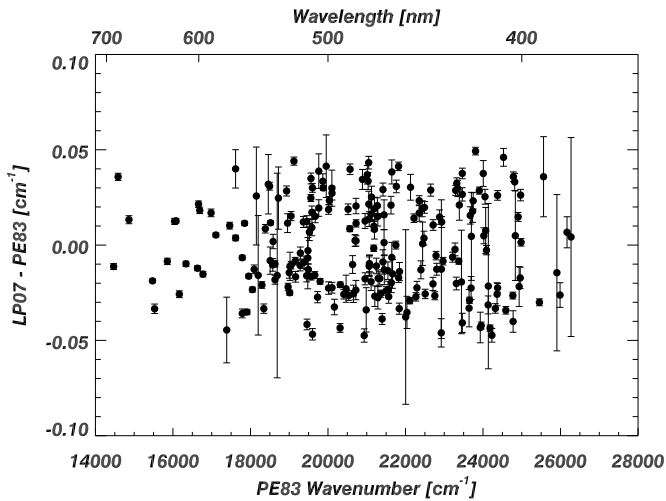}
\caption{Differences between LP07 and PE83 for $231$ lines that were not
correlated by LP07 but that match within $0.05$ cm$^{-1}$.  The uncertainties
of each are the sum in quadrature of the uncertainties from LP07 and PE83.}
\label{figure7}
\end{figure}

Given that their uncertainties were underestimated, we wondered if some of the
lines not identified in LP07 might actually have been measured in PE83.  The
distribution of absolute differences between the unidentified LP07 lines
(classified as `?' in their line list) and the closest PE83 lines is shown in
Figure~\ref{figure5}.  There is a clear population bump of lines with
differences less than about $0.05$ cm$^{-1}$. To investigate these lines we
plotted a HARPS Th/Ar spectrum and one of the PE83 Th/Ne spectra on the same
scale, and found that $231$ of the `?' lines of LP07 were clearly coincident
within $0.05$ cm$^{-1}$ with lines in the spectra of PE83 as illustrated in
Figure~\ref{figure6}.  Most of these lines are perturbed by blending with
nearby lines in the lower resolution HARPS spectra.  Figure~\ref{figure7} shows
the differences between these LP07 lines and the measurements of the same lines
by PE83.  The uncertainties shown are the sum in quadrature of the published
uncertainties.  The ratio between the size of the deviation and the published
uncertainty is as large as $25$, suggesting that for blended lines the LP07
uncertainties may be under-estimated by more than an order of magnitude.

Many additional lines are seen in the HARPS spectra that are not in the spectra
of PE83.  Among these lines are weak thorium lines, argon lines, lines from
trace contamination of other elements, and possibly artifacts from the optics
of HARPS.  Using calculated Ritz wavelengths, as well as compilations such as
the Atomic Spectra
Database\footnote{\url{http://www.nist.gov/pml/data/asd.cfm}} \citep{ASD4.1.0}
and the MIT Wavelength Tables \citep{1969MIT_Tables}, we have identified over
$1700$ of these lines, including $1622$ thorium lines, $94$ argon lines, and a
few lines of contaminant elements.  Since some argon lines are known to show
shifts of as much as $6\times10^{-3}$ cm$^{-1}$ (tens of m/s), depending on the
lamp pressure \citep{whaling2002argon,2006SPIE.6269E..23L}, these lines should
not be used as calibration lines.  Likewise, contaminant lines and lines that
remain unidentified and may be optical artifacts are not suitable for
calibration.  \citet{2006SPIE.6269E..23L} mentions future plans to compare the
ThAr lamp to other lamps (such as thorium-neon) to help identify thorium lines,
but it does not appear that these results were published.

To identify possible contaminant sources, we referred to tables III and IV of
the MIT Wavelength Tables \citep{1969MIT_Tables}.  These lists of sensitive
lines are compiled from empirical observation, and include the lines from the
various elements that are most likely to appear in arc and spark discharges
when only a small quantity of that element is present.  Based on the presence
of these sensitive lines in the HARPS Th/Ar spectra, we have identified lines
of calcium (Ca), chromium (Cr), magnesium (Mg), and manganese (Mn).  PE83 noted
contamination from all of these elements except Mn in their spectra.

\section{Optimized Energy Levels and Ritz Wavelengths}
\label{sec:ritz}

From our own observations and previously published sources we have assembled a
list of \nlines classified thorium lines between $250$~nm and $5500$~ nm that
were used to re-optimize the previously reported levels of Th~I, Th~II, and
Th~III.  Lines were drawn from eight different sources.  Five of these sources
--- PE83, EHW03, LP07, KNS08, and RNS13 --- have already been discussed.  We
also incorporated data from three other publications.
\citet{giacchetti1974proposed} (hereafter, GBCZ74) report FTS observations of
$3100$ thorium lines ($2300$ classified) in the region $900$ nm to $3000$ nm.
\citet{zalubas1974energy} (hereafter, ZC74) and \citet{1976Zalubas} (hereafter,
Z76) include $6485$ classified lines of Th~II and $9587$ classified lines of
Th~I respectively.  ZC74 and Z76 are compilations of measurements from a
variety of experimental sources.

Line classifications from the literature sources were not used directly;
instead, we classified lines from these lists by comparison to Ritz wavelengths
calculated from energy levels given in the actinides
database\footnote{\url{http://web2.lac.u-psud.fr/lac/Database/Tab-energy/Thorium/Th-el-dir.html}}.
In assigning the line identifications to be used for the level optimization we
considered all experimental lines that fell within a fixed search window of
each Ritz prediction.  If a line was observed in more than one source, only the
measurement with the lowest uncertainty was included in the level optimization
and final line list.  A summary of the results obtained by applying this
identification algorithm to the eight experimental lists is presented in
Table~\ref{table3}.

\begin{table}
\caption{Classification Match Summary by Source}
\begin{center}
\begin{tabular}{llll}
\hline \hline
Source	&	Search		&	Classifiable&	Contribution to	\\
		&	Window [cm$^{-1}$]&	Lines		&	Final List		\\
\hline
              GBCZ74	   &$	      0.01  $&$	      2715  $&$	       133 $	\\
   ZC74 $\le 10~000$	   &$	     0.012  $&$	       396  $&$	         6 $	\\
     ZC74 $> 10~000$	   &$	     0.042  $&$	      4784  $&$	      2166 $	\\
    Z76 $\le 10~800$	   &$	     0.012  $&$	      1789  $&$	        55 $	\\
      Z76 $> 10~800$	   &$	      0.05  $&$	      6585  $&$	      2533 $	\\
                PE83	   &$	      0.01  $&$	      9850  $&$	      8644 $	\\
               EHW03	   &$	     0.003  $&$	      3689  $&$	      3606 $	\\
                LP07	   &$	       0.1  $&$	      1622  $&$	      1433 $	\\
               KNS08	   &$	     0.005  $&$	      1685  $&$	       496 $	\\
               RNS13	   &$	     0.003  $&$	      1802  $&$	      1035 $	\\

\hline
\end{tabular}
\end{center}
A summary of how many lines we were able to classify from each of the eight
sources.  If a line appeared in more than one source, only the measurement with
the lowest uncertainty was included in the final line list.  The final column
is the number of lines from each list that were included in the level
optimization and published line list.  \nlines of these lines matched only a
single classification, while \nambig of the lines from PE83, EHW03, KNS08, or
RNS13 matched more than one classification, or a noble gas line and a thorium
Ritz wavenumber.  Another \nblend lines from PE83, EHW03, KNW08, or RNS13
deviate from their Ritz wavenumbers by $5$ or more times their measured
uncertainties.
\label{table3}
\end{table}

\begin{figure}[htbp]
\includegraphics[width=0.5\textwidth]{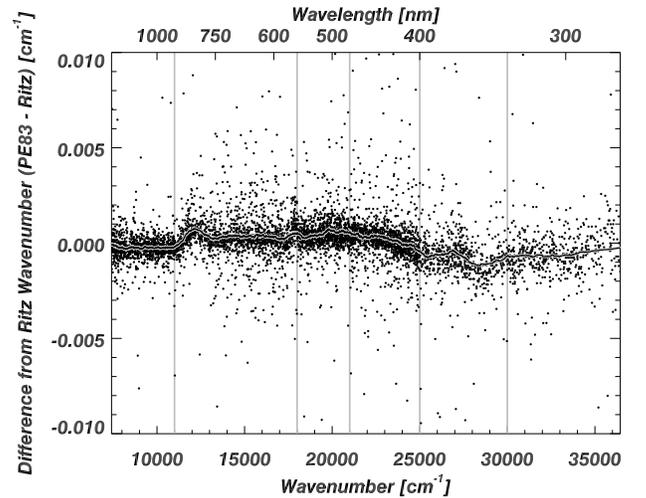}
\caption{The difference between the PE83 wavenumbers and the closest Ritz
wavenumbers, as a function of wavenumber.  We used a moving average with a
window of $200$ points to estimate the systematic errors in the PE83 line list.
Only PE83 wavenumbers with a revised uncertainty (using the coefficients of
Table~\ref{table2} and Equation~\ref{eqn:unc}) of less than $0.0020$ cm$^{-1}$
were used to compute this trend.  Data with larger uncertainties can be used,
but then the window size should be increased, degrading the resolution of the
trend.  The vertical grey lines indicate the separation between the different
regions of the PE83 FTS measurements.  This moving average was subtracted from
the PE83 measurements before optimizing the Th energy levels.}
\label{figure8}
\end{figure}

Prior to the level optimization we modified the wavenumber and uncertainties
from some of the literature sources, as described below:

\begin{enumerate}
\item{{\bf GBCZ74:} The uncertainties of the individual lines are not reported,
but they report an average difference between their measured wavenumbers and
calculated Ritz wavenumbers of $0.0016$ cm$^{-1}$.  We find that the standard
deviation between the published wavenumbers and the Ritz wavenumbers is
$0.0034$ cm$^{-1}$, and it is this value that we assigned as an uncertainty for
all lines in GBCZ74.}
\item{{\bf ZC74, Z76:} The lines reported in ZC74 and Z76 both come from
several sources, and thus it is not too surprising that the uncertainties and
systematic errors vary throughout these lists.  In particular, the
uncertainties are much lower for lines in the near-infrared, all of which come
from GBCZ74.  The standard deviation between the published wavenumber and the
Ritz wavenumbers is $0.0037$ cm$^{-1}$ for ZC74 lines $> 10\,000$ cm$^{-1}$
($1000$ nm), and $0.0026$ cm$^{-1}$ for Z76 lines $> 10\,800$ cm$^{-1}$ ($925$
nm).  At visible wavelengths, the standard deviations are $0.014$ cm$^{-1}$ for
ZC74, and $0.01$ cm$^{-1}$ for Z76.  We have used these values for the
uncertainties of lines in each of these spectral regions.  In most of the
near-UV ($\sigma > 35\,000$ cm$^{-1}$, $\lambda < 300$ nm), the ZC74 wavenumber
uncertainties were too large to be of use.  We also found that the Z76
wavenumbers below $10\,800$ cm$^{-1}$ were systematically low by $0.0025$
cm$^{-1}$ compared to Ritz wavenumbers, which we corrected prior to energy
level optimization.  These lines with their large uncertainties have less
influence on the energy level optimization than more recent FTS measurements,
but they serve to highlight which lines of neutral thorium have been observed
in the lab, and might be seen in other spectra.}
\item{{\bf PE83:} For the measurements of PE83, we first removed the systematic
trends using a moving average, shown in Figure~\ref{figure8}.  The nearest
$200$ points were used to compute the moving average at each point in a plot of
wavenumber difference versus Ritz wavenumber.  The vertical dash-dotted lines
indicate the different sections of the PE83 FTS measurements.  It is worth
noting that these same trends can be detected in figure $1$a of PE83 where they
compare their measured values to their own level optimization results.  In
addition to removing these trends, we have also adjusted the PE83 published
wavenumbers by a factor of $\kappa = 0.22\times10^{-7}$ (see
Fig.~\ref{figure3}) and recalculated their uncertainties as discussed above.}
\item{{\bf EHW03:} We did not modify the wavenumbers of these measurements.
The wavenumber uncertainties of these lines were described as being less than
$0.001$ cm$^{-1}$ for the brightest lines, but no individual uncertainties, nor
the calibration uncertainty, were published.  In order to estimate the
calibration uncertainty, we downloaded the spectrum from the NSO data
archive\footnote{\url{http://diglib.nso.edu}} and redetermined the wavenumber
correction factor using the argon standards of \citet{whaling2002argon} with
the corrections of \citet{sansonetti2007comment}.  We found the wavenumber
correction factor to be $(-7.09\pm0.05)\times10^{-7}$, which is consistent with
the published wavenumber correction factor of $-7.06\times10^{-7}$.  Based on a
comparison to the Ritz wavelengths, we estimated the uncertainties using
Equation~\ref{eqn:unc2}, where $c$ is $0.00107$ and $d$ is $0.242$.}
\item{{\bf LP07:} A comparison between the LP07 lines and the corresponding
PE83 lines suggests that the uncertainties of LP07 were underestimated by a
factor of almost $7$.  We estimated the uncertainties of LP07 using
Equation~\ref{eqn:unc2}, with $c = 0.0796$ and $d = 0.310$ (and no wavenumber
correction factor).  Only LP07 lines that were not found by PE83 were included
in the level optimization.}
\item{{\bf KNS08:} Neither the wavenumbers nor the wavenumber uncertainties of
these measurements were modified.}
\item{{\bf New Measurements (RNS13):}  Neither the wavenumbers nor the
wavenumber uncertainties presented in Tab.~\ref{table1} were modified.}
\end{enumerate}

The classified line lists for Th~I, Th~II, and Th~III, including these
modifications, were taken as input to the program LOPT
\citep{2011CoPhC.182..419K}.  This program performs a least-squares level
optimization to produce improved level values and uncertainties.  It also
calculates Ritz wavelengths based on the new level values with uncertainties
based on the variance and covariance of the combining levels.

Having established improved values for the known levels, we searched for new
energy levels using the unclassified thorium lines from PE83, EHW03, KNS08, and
RNS13.  The known energy levels were taken in groups that included levels of a
single parity and ionization stage with three adjacent J values.  The unknown
lines were summed and differenced with the known levels, producing a large list
of potential new level values.  This list was sorted by energy and searched for
coincident values produced by multiple combinations of lines and levels that
might signal a new level with the opposite parity and central J value.

Because of the high density of potential level values, the likelihood of
accidental coincidences is large.  To assess the validity of accepting a new
level based on a particular number of coincident predictions, we ran Monte
Carlo simulations to determine the number of coincident sums and differences
that would be expected to occur by chance.  For each ionization stage and
combination of adjacent J values, we performed $200$ simulations, adding a
uniformly-distributed random amount between $-0.1$ cm$^{-1}$ and $0.1$
cm$^{-1}$ to the unidentified lines.  Only new energy levels with a probability
of $< 0.01$ of being coincidences were included.  Based on these results we
have accepted \newThI new Th~I energy levels and \newThII new Th~II energy
levels.  Details of the new levels are given in Table~\ref{table4}.  We were
not able to constrain the J value of several of these lines; in these cases,
the two or three possible J values are provided.

Using compilations of neon lines from \citet{saloman2004wavelengths} and
\citet{kramida2006ne}, we found \nneon PE83 lines that were better matches to
neon lines than to thorium Ritz wavelengths.  In all cases, these lines have
larger line widths than the nearby thorium lines, as would be expected for the
gas lines.  PE83 only reported their line widths in a machine-readable table,
which is not publicly available.  This table also includes many faint lines
that were not included in their published atlas.  The PE83 measurements for
lines we have identified as neon are usually more accurate than any previously
reported measurements in the literature.  They are presented in
Table~\ref{table5}.

Our final list of classified thorium lines is provided in Table~\ref{table6}.
This list includes \nambig multiply-classified lines and \nblend lines from
PE83, EHW03, KNS08, and RNS13 with large discrepancies between their measured
and Ritz wavenumbers that were not used in the level optimization.  The first
column is the Ritz wavenumber, which can be used for wavelength calibration and
will usually be more accurate than the measured wavenumber.  The second column
is the Ritz wavenumber uncertainty as determined by the variance and covariance
of the level optimization in LOPT.  The third and fourth columns are the
corresponding Ritz wavelength and uncertainty.  The fifth column is the thorium
ionization stage.  The sixth column is an estimate of the relative line
intensities (described in more detail below).  Columns seven through ten
provide the classification of the line.  The eleventh and twelfth columns are
the measured wavenumber and measured wavenumber uncertainties, modified as
discussed in \S~\ref{ssec:comparison_pe83}.   The second-to-last column
provides the source for the measured wavenumber.  An `A' in the final column
indicates that the line matches more than one thorium Ritz wavenumber; an `AN'
in the final column indicates that the line matches at least one thorium Ritz
wavenumber and one neon wavenumber.  A `B?' in the final column indicates that
the difference between the measured and Ritz wavenumber is more than $5$ times
the measured uncertainty of the line, and therefore was not included in the
level optimization.

In order to help guide users of this line list, we have estimated the relative
intensities of the lines by matching lines from the different line lists and
adjusting their intensities so that they are all on approximately the same
scale.  Placing all line lists on the same intensity scale is complicated by
the fact that the ratio between line intensities from different line lists vary
as a function of wavenumber.  It is also important to note that different lamp
currents could greatly alter the relative intensities of lines from the values
published here \citep[e.g.,][]{2006SPIE.6269E..83K}.  We estimate that these
intensities are only good to an order of magnitude.

A subset of our re-optimized energy levels for Th~I, II, and III are presented
in Table~\ref{table7}.  The level uncertainties are with respect to the ground
state as determined by LOPT.  The first column is the energy of the level, and
the second is its uncertainty.  The third column provides the J value of the
level, and the fourth column indicates the number of measured spectral lines
that were used to determine the energy level.  These energy level uncertainties
are in general an order-of-magnitude better than previously reported level
uncertainties.

\subsection{Implications for Astronomical Measurements}


Our new line list contains the most accurate wavelengths of the thorium
spectrum measured to date.  There are two major astrophysical applications that
will benefit from this line list: precise radial velocity measurements and
measurements of the fine-structure constant as a function of red shift.  

Precise radial velocity measurements (such as those made with the HARPS
spectrograph) rely on relative wavelength shifts.  More accurate wavelengths
can only improve the measured dispersion solution, but the size of the
improvement will depend on the resolution of the spectrometer, the correction
of systematic errors, and the instrumental line profile.  Our line list will
provide the most significant improvement to the dispersion solutions of
high-resolution instruments that were previously utilizing the thorium line
list of LP07.  We note that these improved wavelengths should help characterize
the systematic errors of CCDs and infrared arrays, such as those discovered on
the HARPS spectrograph by \citet{2010MNRAS.405L..16W}.  

On the other hand, measurements of the fine-structure constant ($\alpha$) rely
upon absolute thorium and argon wavelengths to accurately measure slight
changes in the value of $\alpha$ as a function of redshift
\citep{2003MNRAS.345..609M}.  Systematic changes in the thorium wavelengths can
significantly alter the significance of the measured result.  This is
demonstrated very nicely by figure $11$ of \citet{2007MNRAS.378..221M}, which
shows an example of how changes in the thorium wavelengths can transform a
published null result into a possible detection at the level of $2.5$ sigma.
We note that \citet{2007MNRAS.378..221M} utilizes LP07 thorium wavelengths to
solve the dispersion solution of the Very Large Telescope Ultraviolet and
Visual Echelle Spectrograph (UVES).  It is not easy for us to predict how the
application of our line list will affect the measured changes in $\alpha$,
since such an application depends on the exact lines used to solve the
dispersion solution and their proximity to the red-shifted quasar absorption
lines.


\section{Conclusions}
\label{sec:conclusions}

We have compiled a new thorium line list based on the values from seven
previously published sources and our new measurements from the 2-m FTS at NIST.
Using these data, we re-optimized the energy levels of neutral, singly-, and
doubly-ionized thorium and calculated a list of \nlines Ritz wavelengths
between $250$ nm and $5500$ nm.  We included in the list only lines that have
been observed experimentally.  An additional \nambig lines have ambiguous
classifications and another \nblend lines marginally matched a Ritz wavenumber,
but have been included in the list for completeness.  We consider the Ritz
wavelengths from this list to be suitable for use as calibration standards for
new and archival thorium spectra recorded with high-resolution spectrometers.

\section{Acknowledgements}

This research was performed while S. Redman held a National Research Council Research Associateship Award at NIST.

\bibliographystyle{apj}

\begin{table*}
\caption{New thorium energy levels}
\centering
\begin{tabular}{| llcc | llcc | llcc |}
\hline
\multicolumn{12}{|c|}{Even Levels of Thorium-I} \\
\hline
Energy	&	Energy	&	J	&	N	&	Energy	&	Energy	&	J	&	N	&	Energy	&	Energy	&	J	&	N	\\
 	&	Uncertainty	&	 	&		&	 	&	Uncertainty	&	 	&		&	 	&	Uncertainty	&	 	&		\\
(cm$^{-1}$)	&	(cm$^{-1}$)	&	 	&		&	(cm$^{-1}$)	&	(cm$^{-1}$)	&	 	&		&	(cm$^{-1}$)	&	(cm$^{-1}$)	&	 	&		\\
\hline
32605.50954	&	0.0004	&	1	&	16	&	40697.4116	&	0.0013	&	2,3	&	9	&	42993.0860	&	0.0007	&	6	&	8\\
34088.0343	&	0.0005	&	2	&	21	&	40789.3803	&	0.0006	&	5	&	14	&	43067.1989	&	0.0006	&	4	&	10\\
35108.8155	&	0.0006	&	2	&	18	&	41010.1262	&	0.0011	&	3	&	11	&	43090.4322	&	0.0011	&	4	&	15\\
35497.6156	&	0.0009	&	2	&	12	&	41039.8704	&	0.0013	&	2	&	12	&	43155.6660	&	0.0009	&	5,6	&	10\\
37126.4931	&	0.0009	&	1	&	5	&	41285.4358	&	0.0015	&	3	&	16	&	43237.5285	&	0.0014	&	6	&	10\\
37472.7204	&	0.0009	&	1	&	7	&	41500.6175	&	0.0007	&	3	&	11	&	43404.1631	&	0.0017	&	4	&	11\\
38443.8055	&	0.0007	&	3	&	14	&	41548.5831	&	0.0004	&	6	&	9	&	43705.6892	&	0.0009	&	6	&	8\\
38763.4763	&	0.0005	&	5	&	13	&	41557.8339	&	0.0011	&	3	&	12	&	43736.9808	&	0.0006	&	6,7	&	6\\
38960.6542	&	0.0012	&	2	&	13	&	41939.3407	&	0.0005	&	5	&	12	&	43971.4618	&	0.0006	&	7	&	6\\
39160.8499	&	0.0005	&	6	&	10	&	41940.6560	&	0.0013	&	4	&	9	&	44035.7349	&	0.0016	&	4	&	10\\
39419.3351	&	0.0015	&	1	&	9	&	42096.1941	&	0.0015	&	3	&	12	&	44379.0767	&	0.0016	&	5	&	10\\
39964.6289	&	0.0014	&	2	&	13	&	42228.6417	&	0.0012	&	4	&	12	&	44401.9252	&	0.0010	&	6,7	&	6\\
39997.4130	&	0.0005	&	6	&	12	&	42362.6439	&	0.0008	&	4	&	13	&	44663.2790	&	0.0010	&	6,7	&	6\\
39997.6156	&	0.0012	&	3	&	9	&	42366.0020	&	0.0006	&	6	&	7	&	44707.1216	&	0.0014	&	6	&	7\\
40024.9810	&	0.0006	&	2	&	14	&	42396.2783	&	0.0014	&	5	&	13	&	44767.5178	&	0.0009	&	5,6,7	&	7\\
40084.7264	&	0.0010	&	3	&	15	&	42651.3045	&	0.0013	&	3	&	15	&	44823.8521	&	0.0014	&	5,6	&	11\\
40185.1371	&	0.0019	&	2	&	13	&	42760.6053	&	0.0005	&	5	&	14	&	44858.6036	&	0.0009	&	6	&	8\\
40280.9864	&	0.0008	&	2,3	&	14	&	42774.0660	&	0.0011	&	6	&	7	&	45152.6440	&	0.0009	&	6,7	&	7\\
40350.5849	&	0.0012	&	3	&	12	&	42824.6699	&	0.0012	&	4,5	&	12	&	45178.9900	&	0.0011	&	6	&	9\\
40577.1425	&	0.0004	&	3	&	17	&	42839.5573	&	0.0012	&	3	&	11	&	45501.7637	&	0.0017	&	6,7	&	6\\
40577.4147	&	0.0006	&	2	&	13	&	42886.6475	&	0.0012	&	4,5	&	12	&	47289.2212	&	0.0011	&	7	&	6\\
40637.5735	&	0.0006	&	6	&	11	&	42891.6858	&	0.0013	&	4	&	12	&	48292.0919	&	0.0014	&	7	&	5\\
\hline
\multicolumn{8}{|c|}{Odd Levels of Thorium-I}	&	\multicolumn{4}{c|}{Even Levels of Thorium-II} \\
\hline
Energy	&	Energy	&	J	&	N	&	Energy	&	Energy	&	J	&	N	&	Energy	&	Energy	&	J	&	N	\\
 	&	Uncertainty	&	 	&		&	 	&	Uncertainty	&	 	&		&	 	&	Uncertainty	&	 	&		\\
(cm$^{-1}$)	&	(cm$^{-1}$)	&	 	&		&	(cm$^{-1}$)	&	(cm$^{-1}$)	&	 	&		&	(cm$^{-1}$)	&	(cm$^{-1}$)	&	 	&		\\
\hline
25806.5369	&	0.0005	&	1,2	&	11	&	41041.0343	&	0.0008	&	4	&	11	&	23346.8901	&	0.0004	&	1/2,3/2	&	8\\
34013.9166	&	0.0004	&	5,6	&	12	&	41361.1040	&	0.0006	&	4	&	9	&	55351.6949	&	0.0019	&	3/2	&	10\\
35117.9541	&	0.0004	&	5,6	&	9	&	41403.0150	&	0.0005	&	2	&	9	&	56639.9928	&	0.0013	&	7/2	&	7\\
38467.2068	&	0.0007	&	2	&	9	&	41529.3210	&	0.0011	&	2	&	9	&	56717.6330	&	0.0016	&	3/2	&	7\\
38932.6334	&	0.0016	&	2	&	8	&	41585.5483	&	0.0014	&	2	&	10	&	57128.6422	&	0.0012	&	3/2	&	6\\
39243.1896	&	0.0004	&	2	&	9	&	41745.4442	&	0.0006	&	4	&	11	&	63557.6834	&	0.0016	&	7/2	&	7\\
39506.7468	&	0.0006	&	1,2	&	8	&	42003.4633	&	0.0012	&	3	&	8	&	65753.7835	&	0.0012	&	7/2,9/2	&	6\\
39640.1103	&	0.0011	&	2	&	10	&	42237.8531	&	0.0008	&	3	&	8	&	66427.1381	&	0.0017	&	7/2	&	7\\ \cline{9-12}
39887.8122	&	0.0007	&	2	&	8	&	42265.6466	&	0.0012	&	4	&	9	&	\multicolumn{4}{c|}{Odd Levels of Thorium-II} \\ \cline{9-12}
40027.2324	&	0.0007	&	3	&	9	&	42613.6123	&	0.0014	&	2	&	7	&	63620.3291	&	0.0012	&	7/2,9/2,11/2	&	8\\
40100.1245	&	0.0009	&	3	&	14	&	42901.2523	&	0.0008	&	3	&	8	&		&		&		&	 \\
40300.0712	&	0.0006	&	4	&	11	&	43487.4820	&	0.0017	&	4	&	12	&		&		&		&	 \\
40491.3398	&	0.0005	&	3	&	10	&	44153.5692	&	0.0005	&	2	&	8	&		&		&		&	 \\
40933.9985	&	0.0004	&	3,4	&	9	&		&		&		&		&		&		&		&	 \\
\hline

\end{tabular}

New Th~I and Th~II energy levels, as determined from coincident sums between the known energy levels and the unclassified lines of Th.  N is the number of lines used to determine each level.
\label{table4}
\end{table*}

\begin{table*}
\caption{Newly identified neon lines from PE83}
\centering
\begin{tabular}{ll ll ll r c l c}
\hline \hline
            PE83  &           PE83  &  Observed Neon  &  Observed Neon  &           Ritz  &           Ritz &      Source(s)&  Neon Ion& Note \\
      Wavenumber  &    Uncertainty  &     Wavenumber  &    Uncertainty  &     Wavenumber  &    Uncertainty &               &          &      \\
     (cm$^{-1}$)  &    (cm$^{-1}$)  &    (cm$^{-1}$)  &    (cm$^{-1}$)  &    (cm$^{-1}$)  &    (cm$^{-1}$) &               &          &      \\    \hline
$     22428.5580$&$         0.0024$&$       22428.56$&$           0.09$&$      22428.553$&$          0.003$&           KN06&     Ne II&      \\
$      22557.804$&$          0.003$&$       22557.81$&$           0.05$&$       22557.82$&$           0.03$&      P71, KN06&     Ne II&      \\
$     22634.7233$&$         0.0025$&$       22634.70$&$           0.10$&$      22634.690$&$          0.003$&           KN06&     Ne II&    o \\
$     22656.0666$&$         0.0018$&$       22656.06$&$           0.03$&$      22656.059$&$          0.003$&      P71, KN06&     Ne II&      \\
$      22827.765$&$          0.003$&$       22827.76$&$           0.10$&$      22827.770$&$          0.003$&      P71, KN06&     Ne II&      \\
$     22899.1590$&$         0.0024$&$       22899.16$&$           0.05$&$      22899.167$&$          0.003$&      P71, KN06&     Ne II&      \\
$      23001.943$&$          0.003$&$      23001.933$&$          0.015$&$      23001.943$&$          0.016$&      P71, KN06&     Ne II&      \\
$      23300.195$&$          0.003$&$      23300.197$&$          0.015$&$      23300.192$&$          0.016$&      P71, KN06&     Ne II&      \\
$     23301.4339$&$         0.0024$&$      23301.436$&$          0.015$&$      23301.435$&$          0.016$&      P71, KN06&     Ne II&      \\
$      24003.952$&$          0.003$&$     24003.9440$&$         0.0020$&$               $&$               $&     EHR, SBS04&      Ne I&    o \\

\hline
\end{tabular}

A subset of PE83 wavenumbers that correspond to neon lines.  Table~\ref{table5} is published in its entirety in the electronic edition of ApJSS.  Revisions to the wavenumbers and uncertainties of the PE83 lines are described in the text.  Literature values for the Ne~I lines are taken from the compilation of \citet{saloman2004wavelengths}, while those of Ne~II are from \citet{kramida2006ne}.  The source identifications are as follows: 
MH2~=~\citet{meggers1934bs}, 
BAL~=~\citet{burns1950interference}, 
EHR~=~\citet{1970PhDT........32E}, 
P71~=~\citet{1971PhyS....3..133P}, 
PE83~=~\citet{1983ats..book.....P}, 
SBS04~=~\citet{sansonetti2004high}, and KN06~=~\citet{kramida2006ne}.
The last column indicates whether the line matches the observed wavenumber (o) or the ritz wavenumber (r).  Lines without a comment in this column match both.
\label{table5}
\end{table*}

\begin{sidewaystable}
\caption{Thorium line list}
\centering
\begin{tabular}{llllclccccllcc}
\hline \hline
Ritz					&	Ritz Wavenumber			&	Ritz Vacuum		&	Ritz Wavelength		&	Species	&	Relative	&	Odd			&	Odd		&	Even			&	Even		&	Measured		&	Measured				&	Source	&	Note	\\
Wavenumber			&	Uncertainty				&	Wavelength		&	Uncertainty			&			&	Intensity	&	Energy		&	J		&	Energy		&	J		&	Wavenumber	&	Wavenumber			&			&		\\
($\sigma$) [cm$^{-1}$]	&	($\delta_\sigma$) [cm$^{-1}$]	&	($\lambda$) [\AA]	&	($\delta_\lambda$) [\AA]	&			&			&	[cm$^{-1}$]	&			&	[cm$^{-1}$]	&			&	[cm$^{-1}$]	&	Uncertainty [cm$^{-1}$]	&			&		\\
\hline
   $9816.11445$    & $0.00010$    &  $10187.33028$    & $0.00011$    &      Th I    &     $180$    &   $23113$    &       $4$    &   $13297$    &       $4$    &    $9816.1146$    &       $0.0004$    &      PE83    &        \\
    $9819.6098$    &  $0.0003$    &   $10183.7040$    &  $0.0003$    &     Th II    &       $0$    &   $17837$    &     $1/2$    &   $08018$    &     $3/2$    &     $9819.616$    &        $0.003$    &    GBCZ74    &        \\
   $9819.91929$    & $0.00006$    &  $10183.38309$    & $0.00006$    &      Th I    &     $290$    &   $17224$    &       $2$    &   $27044$    &       $3$    &    $9819.9191$    &       $0.0003$    &      PE83    &        \\
   $9821.92006$    & $0.00005$    &  $10181.30868$    & $0.00005$    &      Th I    &    $1500$    &   $22669$    &       $3$    &   $12847$    &       $3$    &   $9821.92015$    &      $0.00011$    &      PE83    &        \\
   $9823.30548$    & $0.00020$    &  $10179.87277$    & $0.00020$    &      Th I    &      $37$    &   $28372$    &       $1$    &   $18549$    &       $2$    &    $9823.3059$    &       $0.0011$    &      PE83    &        \\
   $9825.30957$    & $0.00006$    &  $10177.79636$    & $0.00006$    &      Th I    &    $1200$    &   $14206$    &       $4$    &   $24032$    &       $4$    &   $9825.30956$    &      $0.00012$    &      PE83    &        \\
    $9830.8247$    &  $0.0004$    &   $10172.0866$    &  $0.0004$    &      Th I    &      $59$    &   $28932$    &       $4$    &   $38763$    &       $5$    &   $ 9830.8230$    &       $0.0008$    &      PE83    &    A   \\
   $9830.82333$    & $0.00019$    &  $10172.08800$    & $0.00020$    &     Th II    &      $59$    &   $20686$    &     $5/2$    &   $10855$    &     $7/2$    &   $ 9830.8230$    &       $0.0008$    &      PE83    &    A   \\
   $9839.51266$    & $0.00010$    &  $10163.10497$    & $0.00010$    &      Th I    &     $380$    &   $25703$    &       $2$    &   $15863$    &       $2$    &     $9839.512$    &        $0.003$    &       Z76    &        \\
   $9843.30405$    & $0.00011$    &  $10159.19040$    & $0.00011$    &      Th I    &      $33$    &   $19039$    &       $2$    &   $28882$    &       $2$    &    $9843.3041$    &       $0.0011$    &      PE83    &        \\
   $9855.08606$    & $0.00003$    &  $10147.04483$    & $0.00003$    &      Th I    &     $720$    &   $16217$    &       $2$    &   $06362$    &       $2$    &   $9855.08586$    &      $0.00016$    &      PE83    &        \\

\hline
\end{tabular}

A sample of the compiled thorium line list.  Table~\ref{table6} is published in
its entirety in the electronic edition of ApJSS.  The energy levels have been
rounded down to an integer.  The actual values of the energy levels can be
found in Table~\ref{table7}.  GBCZ74~=~\citet{giacchetti1974proposed},
ZC74~=~\citet{zalubas1974energy}, Z76~=~\citet{1976Zalubas},
PE83~=~\citet{1983ats..book.....P}, EHW03~=~\citet{2003JQSRT..78....1E},
LP07~=~\citet{2007A&A...468.1115L}, KNS08~=~\citet{kerber2008th}, RNS13~=~This
publication.  An `A' in the last column indicates a measured line that matches
more than one classification.  An `B?' indicates a line that deviates from the
Ritz wavenumber by $5$ or more times its measured uncertainty, and therefore
was not included in the level optimization.
\label{table6}
\end{sidewaystable}

\begin{table*}
\caption{Optimized thorium energy levels}
\centering
\begin{tabular}{| llcc | llcc | llcc |}
\hline
\multicolumn{12}{|c|}{Even Levels of Thorium}		\\
\hline
\multicolumn{4}{|c|}{Th I}										&	\multicolumn{4}{|c|}{Th II}									&	\multicolumn{4}{|c|}{Th III}									\\
\hline
Energy		&	Energy		&	J	&	N	&	Energy		&	Energy		&	J	&	N	&	Energy		&	Energy		&	J	&	N	\\
			&	Uncertainty	&		&		&				&	Uncertainty	&		&		&				&	Uncertainty	&		&		\\
$[$cm$^{-1}$]	&	[cm$^{-1}$]	&		&		&	[cm$^{-1}$]	&	[cm$^{-1}$]	&		&		&	[cm$^{-1}$]	&	[cm$^{-1}$]	&		&		  \\
\hline
    $  0.000000    $    &    $  0.000010    $    &    $         2    $    &    $       189    $    &    $46902.5368    $    &    $    0.0011    $    &    $       5/2    $    &    $        22    $    &    $7176.10661    $    &    $    0.0003    $    &    $         2    $    &    $        10    $   \\
    $2558.05675    $    &    $  0.000018    $    &    $         0    $    &    $        48    $    &    $1521.89632    $    &    $   0.00010    $    &    $       5/2    $    &    $        80    $    &    $  63.26679    $    &    $    0.0003    $    &    $         2    $    &    $         9    $   \\
    $2869.25916    $    &    $  0.000017    $    &    $         3    $    &    $       198    $    &    $1859.93843    $    &    $   0.00011    $    &    $       3/2    $    &    $        60    $    &    $ 7875.8244    $    &    $    0.0004    $    &    $         1    $    &    $         5    $   \\
    $3687.98718    $    &    $  0.000013    $    &    $         2    $    &    $       181    $    &    $4113.35932    $    &    $   0.00011    $    &    $       5/2    $    &    $        75    $    &    $9953.58098    $    &    $    0.0003    $    &    $         3    $    &    $        12    $   \\
    $3865.47377    $    &    $  0.000016    $    &    $         1    $    &    $       126    $    &    $4146.57708    $    &    $   0.00011    $    &    $       7/2    $    &    $        70    $    &    $10440.2372    $    &    $    0.0004    $    &    $         2    $    &    $         6    $   \\
    $4961.65883    $    &    $  0.000019    $    &    $         4    $    &    $       176    $    &    $27631.2250    $    &    $   0.00024    $    &    $       3/2    $    &    $        20    $    &    $10542.8987    $    &    $    0.0003    $    &    $         4    $    &    $         9    $   \\
    $5563.14151    $    &    $  0.000016    $    &    $         1    $    &    $       114    $    &    $27937.0722    $    &    $    0.0003    $    &    $      11/2    $    &    $        17    $    &    $11961.1316    $    &    $    0.0007    $    &    $         0    $    &    $         3    $   \\
    $6362.39598    $    &    $  0.000016    $    &    $         2    $    &    $       183    $    &    $28011.1578    $    &    $    0.0003    $    &    $       3/2    $    &    $        13    $    &    $15148.5193    $    &    $   0.00023    $    &    $         4    $    &    $        10    $   \\
    $7280.12322    $    &    $  0.000023    $    &    $         2    $    &    $       179    $    &    $28026.3485    $    &    $    0.0005    $    &    $       5/2    $    &    $        18    $    &    $16037.6412    $    &    $    0.0003    $    &    $         2    $    &    $         9    $   \\
    $7502.28763    $    &    $  0.000016    $    &    $         3    $    &    $       184    $    &    $28823.6538    $    &    $   0.00025    $    &    $       5/2    $    &    $        23    $    &    $17887.4092    $    &    $    0.0003    $    &    $         5    $    &    $         7    $   \\
    \hline
    \multicolumn{12}{|c|}{Odd Levels of Thorium}\\
    \hline
    $39468.6795    $    &    $    0.0004    $    &    $         3    $    &    $        18    $    &    $37716.3253    $    &    $    0.0012    $    &    $       1/2    $    &    $         4    $    &    $11276.8070    $    &    $    0.0003    $    &    $         5    $    &    $         8    $   \\
    $33270.5454    $    &    $    0.0003    $    &    $         4    $    &    $        20    $    &    $6168.35582    $    &    $   0.00013    $    &    $       7/2    $    &    $        69    $    &    $2527.09550    $    &    $    0.0003    $    &    $         3    $    &    $        11    $   \\
    $8243.60209    $    &    $   0.00003    $    &    $         2    $    &    $       132    $    &    $37756.7895    $    &    $    0.0008    $    &    $       7/2    $    &    $        17    $    &    $ 3181.5024    $    &    $    0.0004    $    &    $         2    $    &    $         7    $   \\
    $37510.1884    $    &    $    0.0007    $    &    $         6    $    &    $         4    $    &    $43382.7887    $    &    $    0.0008    $    &    $       5/2    $    &    $        13    $    &    $3188.30107    $    &    $    0.0003    $    &    $         4    $    &    $        10    $   \\
    $33294.9255    $    &    $    0.0006    $    &    $         3    $    &    $        18    $    &    $6691.38730    $    &    $   0.00013    $    &    $       3/2    $    &    $        56    $    &    $20710.9487    $    &    $    0.0006    $    &    $         1    $    &    $         3    $   \\
    $10414.1370    $    &    $   0.00004    $    &    $         4    $    &    $       143    $    &    $6700.18627    $    &    $   0.00016    $    &    $       9/2    $    &    $        63    $    &    $ 4489.6410    $    &    $    0.0004    $    &    $         5    $    &    $         4    $   \\
    $10526.5444    $    &    $   0.00003    $    &    $         3    $    &    $       177    $    &    $43744.0904    $    &    $    0.0015    $    &    $       1/2    $    &    $         5    $    &    $13208.2137    $    &    $    0.0004    $    &    $         2    $    &    $         6    $   \\
    $10783.1553    $    &    $   0.00007    $    &    $         2    $    &    $       108    $    &    $7331.48541    $    &    $   0.00013    $    &    $       5/2    $    &    $        69    $    &    $4826.82620    $    &    $   0.00024    $    &    $         3    $    &    $        14    $   \\
    $11197.0312    $    &    $   0.00003    $    &    $         5    $    &    $       113    $    &    $63620.3291    $    &    $    0.0012    $    &    $7/2,9/2,11    $    &    $         8    $    &    $5060.54386    $    &    $   0.00023    $    &    $         3    $    &    $        12    $   \\
    $11241.7304    $    &    $   0.00005    $    &    $         3    $    &    $       178    $    &    $37846.1686    $    &    $    0.0005    $    &    $       5/2    $    &    $        20    $    &    $42259.7128    $    &    $    0.0027    $    &    $         0    $    &    $         1    $   \\
    \hline

\end{tabular}

A subset of the compiled Th-I (left), Th-II (center), and Th-III (right) energy
levels.  Table~\ref{table7} is published in its entirety in the electronic
edition of ApJSS.  The even levels are in the top half of the table, and the
odd levels are in the bottom half of the table.  The entire list of optimized
energy levels can be found online.
\label{table7}
\end{table*}

\end{document}